\documentclass{iaus}
\usepackage{graphicx}

\def\mearth{ M_{\oplus}}

\def\rjup{\rm R_{J}}

\def\rsun{\rm R_{\odot}}
\def\rs{R_{\star}}

\def\fdg{\mbox{\ensuremath{.\!\!^\circ}}}
\title[\emph{Spitzer} Transit Observations of HD~149026b] %% give here short title %%
{A Precise Estimate of the Radius of HD~149026b}

\author[Nutzman et al.]   %% give here short author list %%
		{Philip Nutzman$^{1}$,
		David Charbonneau$^{1,2}$,
		Joshua N. Winn$^{3}$,
		Heather A. Knutson$^{1}$,
		Jonathan~J.~Fortney$^{4}$,
		Matthew J. Holman$^{1}$,
		Eric Agol$^{5}$	 
		}

\affiliation{$^1$Harvard-Smithsonian Center for Astrophysics, 60 Garden St., 
Cambridge, MA 02138\\ email: {\tt pnutzman@cfa.harvard.edu} \\[\affilskip]
$^2$Alfred P. Sloan Research Fellow\\ [\affilskip]
$^3$Department of Physics, and Kavli Institute for
Astrophysics and Space Research, Massachusetts Institute of
Technology, Cambridge, MA 02139, USA\\ [\affilskip]
$^4$Department of Astronomy and Astrophysics, UCO/Lick Observatory, University
of California, Santa Cruz, CA 95064 \\ [\affilskip]
$^5$Department of Astronomy, University of Washington, Box 351580, Seattle, WA
98195\\
}

\pubyear{2008}
\volume{xxx}  %% insert here IAU Symposium No.
\pagerange{119--126}

\jname{IAU Symposium No. 253}
\editors{A.C. Editor, B.D. Editor \& C.E. Editor, eds.}
\begin{document}

\maketitle

\begin{abstract}
We present \emph{Spitzer} 8 $\mu$m transit observations of the
extrasolar planet system HD~149026.  At this wavelength, transit light curves are weakly affected by stellar limb-darkening, allowing for a simpler and more accurate determination of planetary parameters.  We measure a planet-star radius ratio of $R_p/\rs = $0.05158$ \pm $0.00077, and in combination with ground-based data and independent constraints on the stellar mass and radius, we derive an orbital inclination of $i = $85$ \fdg $4$ ~^{+0 \fdg9}_{-0 \fdg8} $ and a planet radius of 0.755 $\pm$ 0.040 $~\rjup$.  These measurements further support models in which the planet is greatly enriched in heavy elements. 

\end{abstract}

\firstsection % if your document starts with a section,
\section{Introduction}
The hot, Saturn-mass exoplanet HD 149026b ($M = 114 \mearth$; discovered by Sato et al. 2005) has garnered much attention due to its potential to directly test models of planet formation.  The planet's 
small observed radius for its mass implies that roughly 2$/$3 of its mass is in
the form of heavy elements-- more than found in all of the Solar System planets combined.  
The existence of such a metal-laden planet orbiting a very metal-rich host
star ([Fe/H] $=$ 0.36) is a smoking gun for the core-accretion theory (e.g.,
Pollack et al. 1996), though the immense quantity of heavy elements is still a
challenge to canonical models of core accretion.  HD 149026b has also received
notice for its scorching day-side brightness temperature (2300 K), which is
well in excess of its expected equilibrium temperature.  The planet may be a standard bearer for the proposed `pM' 
class of planets (Fortney et al. 2008), which are characterized by hot
stratospheres and large day/night temperature contrasts caused by opacity at
visible wavelengths due to gaseous TiO and VO high in their atmospheres. 

Unfortunately, the present fractional uncertainty in the key observable
parameter, the planetary radius $R_p$, is $7 \%$ (Winn et al. 2008).  This
study is inspired by the potential of infrared photometry with the
\emph{Spitzer Space Telescope} to reduce this uncertainty.  Because of the
near absence of stellar-limb darkening in the infrared, transit light curve
modeling is greatly simplified and gives results largely independent of
assumptions about limb-darkening coefficients.  We present
\emph{Spitzer} 8 $\mu$m  observations of the transit of HD 149026.  Together
with previously published ground-based data, we derive precise constraints on
the radius of HD~149026b and other system parameters.\footnote{For a detailed description of our analysis, please consult Nutzman et al. (2008).}

\section{Observations and Analysis}

%%%%%%%%%%%%% figure 1 %%%%%%%%%%%
\begin{figure}[b]
%\centering
\begin{center}
\includegraphics[width=0.95\textwidth]{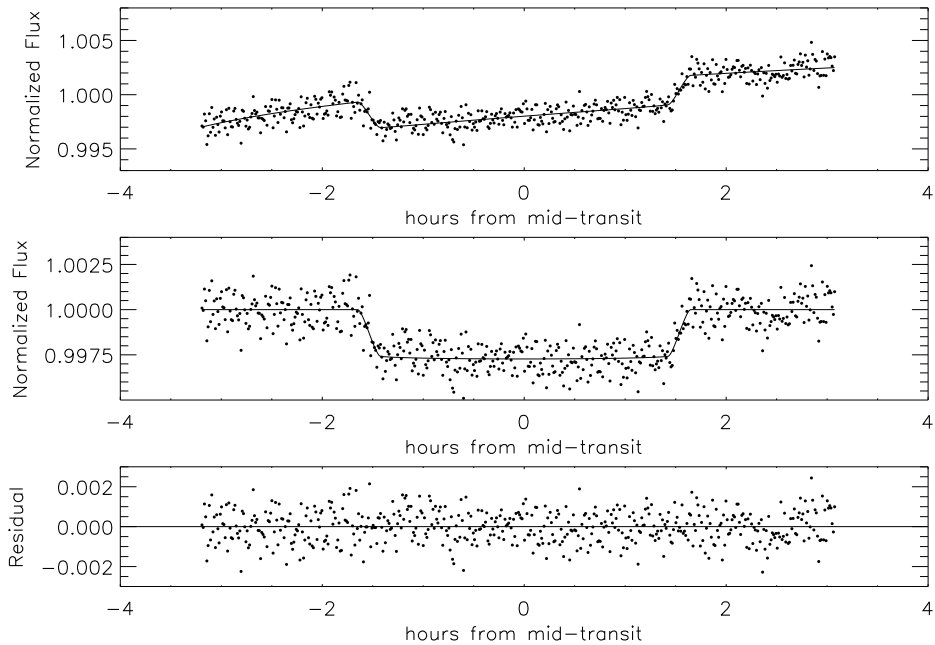}
\caption{\emph{Spitzer} transit photometry for HD 149026, with 40 second
resolution (bins of 100 images).  The top panel displays the raw light curve
and exhibits the well-known IRAC 8 $\mu$m detector ramp (e.g., Knutson et al
2007).  The middle panel displays the ramp-corrected light curve together with the best-fitting transit model.  At bottom are the
residuals from the best-fit.  The root-mean-square residual is only 15 $\%$
greater than the expected photon noise.}
\end{center}
\end{figure}
%%%%%%%%%%%%%%%%%%%%%%%%%%%%%%%%%%%%%%

We observed the UT August 14, 2007 transit of HD 149026 through the IRAC 8$\mu$m channel on the \emph{Spitzer}
space telescope (see Figure 1).  We obtained 67,008 images in IRAC subarray
mode at a 0.4 s cadence.  The data exhibits very low levels of
time-correlated, or ``red'', noise, as demonstrated by Figure 2.  Figure 2
also indicates that, at 15 minute resolution (bins of roughly 2,000 images),
the photometric precision dips below $2 \times 10^{-4}$. 
In order to achieve the strongest possible constraints on the
system parameters, we combined our data with previously published ground-based
photometry (Sato et al. 2005, Charbonneau et al. 2006, Winn et al. 2008).

%and adopted a constraint on the
%stellar mass (from Sato et al. 2005) and on the stellar radius (based on NIR
%magnitudes and the observed parallax). 

We employed Markov Chain Monte Carlo to derive
best-fit parameters and uncertainties (see Table 1).  We
imposed an external constraint on the stellar mass, as derived by Sato et al.
(2005).  We also employed the empirical $V,$ $K$ relations of Kervella et al.
(2004) for the
angular diameters of dwarf stars, and
combined with the \emph{Hipparcos} parallax to derive an
external constraint on the radius of the star HD 149026.

%%%%%%%%%%%%%%%%%%%%%%%%%%%%%%%%%%%%%%
\begin{table}
\begin{center}
\begin{tabular}{|l|l|l|l|}

\hline
\multicolumn{4}{|c|}{Estimates of the HD~149026}\\
\multicolumn{4}{|c|}{ System Parameters}\\
\hline
\small{Parameter} & \small{Median} & \small{15.9$^{th}$ Perc.}&
\small{84.1$^{th}$ Perc.}\\
\hline
$R_p/\rs$	~~~~~	& $0.05147	$~~~~~~& $-0.00077$	~~~~~~~~& $+0.00076$\\
$i\rm{~[deg]}$			& $ 85.3 	$& $-0.8 $	& $+0.9$	\\
$a/\rs$				& $6.20	$	& $-0.25$	& $+0.28	$\\
$R_p [\rjup]$ 			&$ 0.755  	$& $-0.040 $	& $+0.040 	$\\
$\rs  [\rsun]$  	        &$ 1.497   	$& $-0.069$   	& $+0.069 	$\\
\hline
\multicolumn{4}{l}{~}\\
%\multicolumn{4}{l}{\caption{\large{\textbf{\emph{Table 1:}}}Best-fit
%parameters and estimates of HD 149026 system properties derived through a
%Markov Chain Monte Carlo analysis.}}

\end{tabular}
\caption{Best-fit
parameters and uncertainties of HD 149026 system properties derived through a
Markov Chain Monte Carlo analysis.}

\end{center}
\end{table}

IRAC 8$\mu$m observations exhibit a detector ramp (see Figure 1 and e.g., Knutson et al 2007).  Our MCMC
model includes a simultaneous analysis of a ramp model and the transit model,
hence our error bars include the resulting correlations and uncertainties.
The data, with and without the ramp correction and binned 100:1 are shown in
Figure 1, together with the best-fitting transit model.

%%%%%%%%%%%%% figure 2 %%%%%%%%%%%
\begin{figure}[b]
%\centering
\begin{center}
\includegraphics[width=\textwidth]{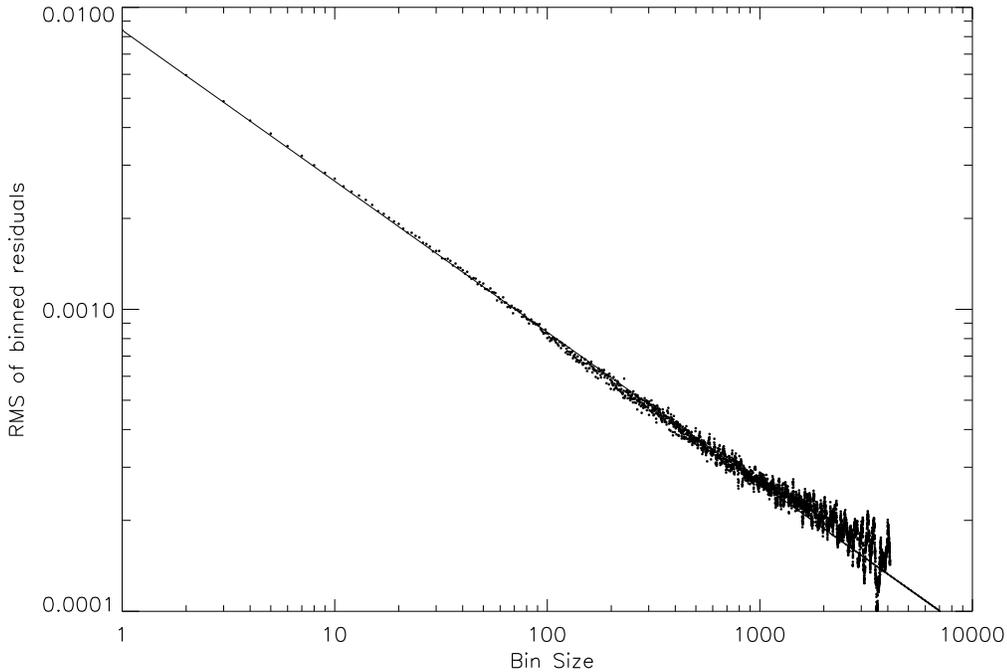}
\caption{The root-mean-square of binned residuals vs. bin size.  The solid line is proportional to $N^{-1/2}$.  }
\end{center}
\end{figure}
%%%%%%%%%%%%%%%%%%%%%%%%%%%%%%%%%%%%%%

%%%%%%%%%%%%% figure 3 %%%%%%%%%%%
\begin{figure}
%\centering
\includegraphics[width=1.0\textwidth]{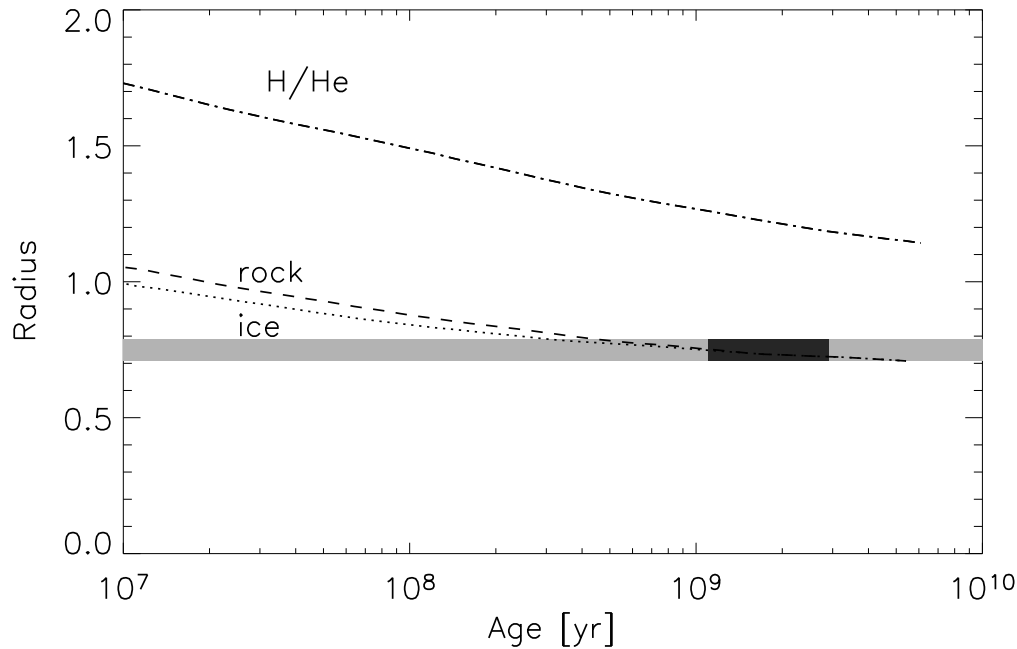}
\caption{The radius of HD 149026b for three planet models of Fortney
et al. (2006).  The light grey band represents the observational radius
constraint from this work. The dark grey box includes the age constraint
of Sato et al. (2005). The ``icy'' model is for an H/He envelope and a $77
\mearth$ core of ice, while the ``rocky'' model is for an He/He envelope and a
$65 \mearth$ core of olivine. A solar composition model (dot-dashed line), in which the planet is predominantly H/He, is excluded by the observational constraints.}
\end{figure}
%%%%%%%%%%%%%%%%%%%%%%%%%%%%%%%%%%%%%%

\section{Discussion}

We have presented and analyzed \emph{Spitzer} 8 $\mu$m transit observations of
the HD 149026 system.  By incorporating previously published data, and
adopting constraints on the stellar mass and radius, we improve the
determination of the planetary radius to $R_p = 0.755 \pm 0.040 ~\rjup$.  

Our measurement reinforces previous findings of the intriguingly small radius of
HD 149026b.  In comparison, models in which HD 149026b is
composed purely of H/He require a radius greater than $1.1 ~\rjup$ (see Figure 3).

~\\
~\\

\end{document}